\documentclass[aps,prb,twocolumn,superscriptaddress,showpacs,preprintnumbers,amsmath,amssymb,floatfix]{revtex4}
\usepackage{graphicx}
\usepackage{dcolumn}
\usepackage{bm}
\usepackage{physics}
\usepackage{xcolor}
\usepackage[latin1]{inputenc}
\graphicspath{{./figs/}}

\begin{document}

\title{Tuning of exciton type by environmental screening}

\author{Igor L. C. Lima}
\affiliation{Departamento de
F\'{\i}sica, Universidade Federal do Cear\'a, Caixa Postal 6030, 60455-760 Fortaleza, Cear\'a, Brazil}
\affiliation{Departamento de F\'isica, Instituto Tecnol\'ogico de Aeron\'autica, DCTA, 12228-900 São Jos\'e dos Campos, Brazil }

\author{M. V. Milo\v{s}evi\'c} 
\affiliation{Department of Physics, University of Antwerp, Groenenborgerlaan 171, B-2020 Antwerp, Belgium}

\author{F. M. Peeters} 
\affiliation{Department of Physics, University of Antwerp, Groenenborgerlaan 171, B-2020 Antwerp, Belgium}
\affiliation{Departamento de
F\'{\i}sica, Universidade Federal do Cear\'a, Caixa Postal 6030, 60455-760 Fortaleza, Cear\'a, Brazil}

\author{Andrey Chaves} \email{andrey@fisica.ufc.br}
\affiliation{Departamento de
F\'{\i}sica, Universidade Federal do Cear\'a, Caixa Postal 6030, 60455-760 Fortaleza, Cear\'a, Brazil}
\affiliation{Department of Physics, University of Antwerp, Groenenborgerlaan 171, B-2020 Antwerp, Belgium}

\begin{abstract}
We theoretically investigate the binding energy and electron-hole (e-h) overlap of excitonic states confined at the interface between two-dimensional materials with type-II band alignment, i.e. with lowest conduction and highest valence band edges placed in different materials, arranged in a side-by-side planar heterostructure. We propose a variational procedure within the effective mass approximation to calculate the exciton ground state and apply our model to a monolayer MoS$_2$/WS$_2$ heterostructure. The role of non-abrupt interfaces between the materials is accounted for in our model by assuming a W$_{x}$Mo$_{1-x}$S$_2$ alloy around the interfacial region. Our results demonstrate that: (i) interface-bound excitons are energetically favorable only for small interface thickness and/or for systems under high dielectric screening by the materials surrounding the monolayer, and that (ii) the interface exciton binding energy and its e-h overlap are controllable by the interface width and dielectric environment.    
\end{abstract}

\maketitle

\section{Introduction}

Two-dimensional (2D) transition metal dichalcogenides (TMDs) are few-layer semiconductors with the general chemical formula MX$_2$, where M is a transition metal atom and X is a chalcogen atom.\cite{kolobov2016two} In monolayer form, this class of 2D materials presents an unique combination of atomic-scale thickness, direct bandgap, and valley-spin physics, \cite{manzeli20172d} which makes them interesting not only for fundamental studies, but also for future applications in optoelectronics and atomically thin photodetectors. \cite{konstantatos2018current}

Engineered semiconductor heterojunctions and heterostructures have been one of the most promising ways of tuning and obtaining desired electronic and transport properties out of pristine materials. \cite{mimura1980new,kroemer1963proposed} In the context of 2D materials, heterostructures based either on side-by-side \cite{duan2014lateral} or stacked \cite{rivera2015observation} TMDs layers have been attracting attention, as an exciting way to obtain spatially separated excitons, where electrons and holes reside in different materials, as a consequence of the so called type-II (staggered) band alignment. Due to its significantly longer lifetime, this kind of exciton is of particular interest in the context of photodetectors and energy absorption applications.\cite{li2021photodetectors} However, despite being widely experimentally probed in several TMD van der Waals stacked heterostructures, \cite{kunstmann2018momentum, nagler2019interlayer, ovesen2019interlayer} the experimental observation of spatially separated excitons in side-by-side TMD heterostructures has been elusive. \cite{zhou2020suppressing}

In this paper, we study the excitonic states in a planar side-by-side monolayer heterostructure with type-II band alignment. The model proposed here, which is based on the effective mass theory and uses a combination of numerical diagonalization and variational methods, is applied to the specific case of a MoS$_2$/WS$_2$ heterostructure, but can be straightforwardly adapted to other combinations of materials, {with the only caveat that the present calculations do not account for mass anisotropy, since neither MoS$_2$ nor WS$_2$ exhibit anisotropic effective masses}. Our results demonstrate that, despite the type-II band alignment, the strong electron-hole (e-h) interaction present in a monolayer surrounded by vacuum overcomes the bands mismatch and leads to an exciton state where electron and hole are situated in the same material, i.e. a spatially direct exciton, as the most energetically favourable state. Considering capping layers with higher dielectric constant, however, helps to suppress the otherwise strong e-h interaction and eventually lead to a spatially indirect interface exciton. Non-abrupt interfaces\cite{shimasaki2022directional,sahoo2018one,huang2014lateral} between the materials are accounted for as a W$_{\chi}$Mo$_{1-\chi}$S$_2$ alloy with a concentration gradient $\chi(x)$ that smoothly increases from 0 to 1 along the coordinate $x$ across the interface between MoS$_2$ and WS$_2$\cite{zhang2018strain}$^,$\cite{lim2014stacking}. The effects of the width of such gradual interface, as well as of the dielectric environment, on the exciton energy and e-h overlap of the spatially indirect excitons are discussed.

\section{Model and methods}

The system is defined as a lateral junction of two different 2D semiconductor materials. For practical purposes, we consider here the specific case of a molybdenum disulfide (MoS$_2$) and tungsten disulfide (WS$_2$) heterostructure, {but changing the materials combination would simply lead to different values for effective masses and band offsets in the model proposed here.}

The interface between the materials in such side-by-side monolayer heterostructure is assumed to be smooth, by considering a W$_{\chi(x)}$Mo$_{1-\chi(x)}$S$_2$ alloy whose concentration $\chi(x)$ varies along an interface region of width $w$, as sketched in Fig. \ref{fig:System}(a).{\cite{shimasaki2022directional,sahoo2018one,huang2014lateral} Notice that, in this case, microscopic aspects of the interface, such as edge types and their role on K-to-K' scattering \cite{tang2021moire, qiu2015nonanalyticity, yu2015valley,wu2015exciton}, are no longer relevant. For the sake of simplicity, we approximate the concentration profile by a linear function}
\begin{equation}
\label{eq.chi}
    \chi(x) = 
    \begin{cases}
        1 &\mbox{if } x < -w/2, \\
        1 - \frac{x+w/2}{w} &\mbox{if } -w/2 < x < w/2, \\
        0 &\mbox{if } x > w/2, \\
    \end{cases}
\end{equation}
which is sketched as a blue solid line in Fig. \ref{fig:System}(b). {Nevertheless, in the theoretical framework proposed here, changing this concentration profile by a more realistic function is straightforward.}

\begin{figure}[h!]
    \centering
    \includegraphics[width = \linewidth]{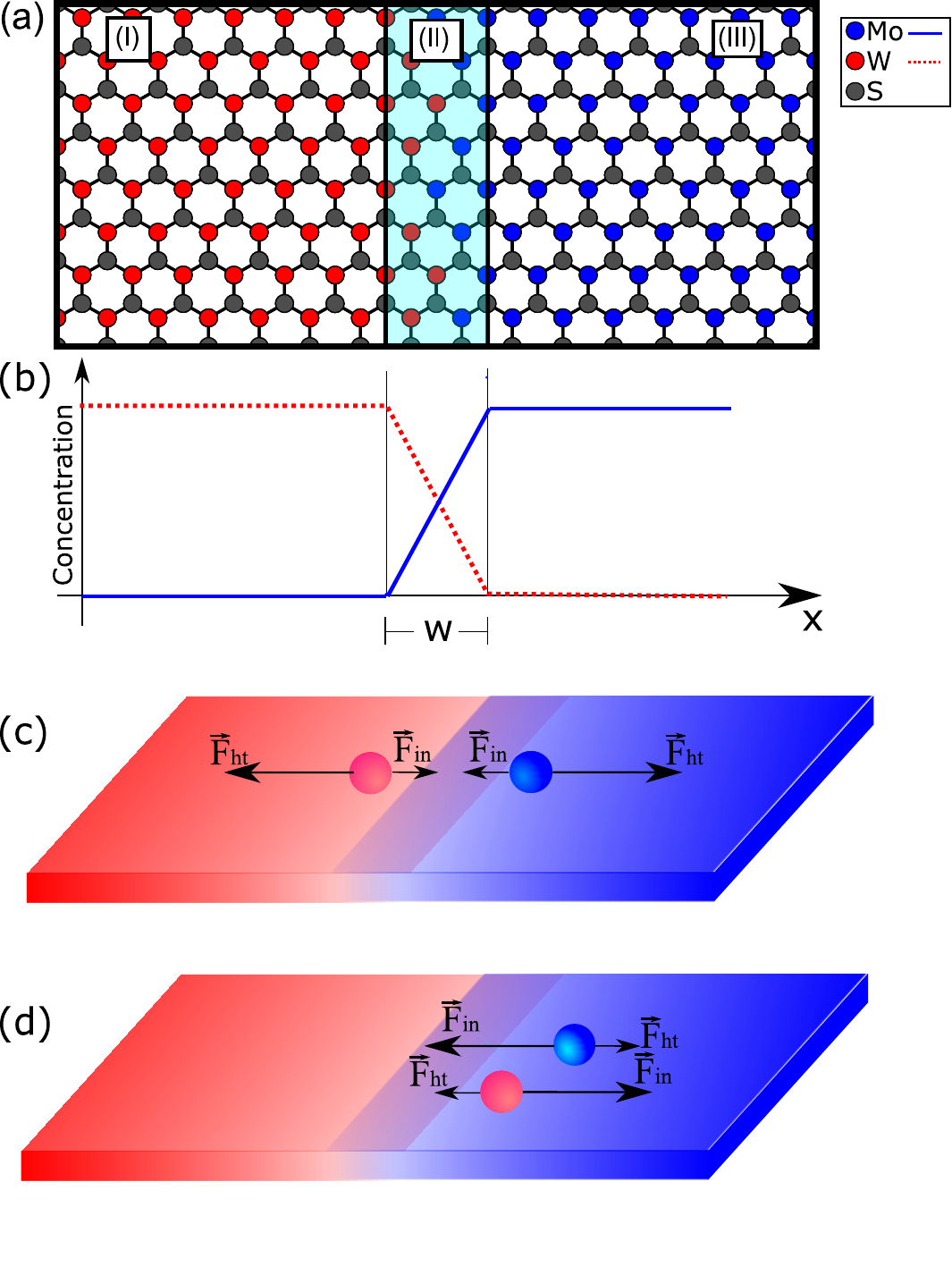}
    \caption{(color online) (a) Sketch of the side-by-side monolayer heterostructure investigated here. From the left to the right side, a W$_{\chi}$Mo$_{1-\chi}$S$_2$ alloy goes from pristine WS$_2$ to pristine MoS$_2$, with a smoothly increasing (decreasing) Mo (W) concentration along the interface region. (b) Sketch of the tungsten concentration function $\chi(x)$ (red dashed), along with the complementary Mo concentration function $1-\chi(x)$ (blue solid) along the interface, with width $w$. (c) Sketch of the interaction mechanisms in an interface exciton, where the electron is in the left side and the hole is in the right side. The heterostructure potential produces a force $\vec{F_{ht}}$ strong enough to keep the electron and the whole separated, while the e-h interaction force ($\vec{F_{in}}$) keeps them close enough to form a bound exciton at the interface. (d) Sketch of the interaction mechanisms in an system in which the e-h attraction force ($\vec{F_{in}}$) is stronger than the heterostructure forces, thus being able to push the whole exciton to the pure MoS$_2$ region.} 
    \color{black}
    \label{fig:Sketch}
\end{figure}

Let $V_{e(h)}(x_{e(h)})$ be the function that represents the heterostructure potential, due to conduction (valence) bands mismatch for the electron (hole). The form of this potential is determined by the tungsten concentration $\chi(x)$, represented by\cite{haastrup2018computational,gjerding2021recent}
\begin{equation}\label{eq.bandoffset}
    \begin{aligned}
        & V_e(x_e) = 353\chi(x_e) - 280\chi(x_e)(1-\chi(x_e)),\\
        & V_h(x_h) = 208(1-\chi(x_h)),
    \end{aligned}
\end{equation}
in units of meV, where a positive effective mass is already defined for the hole. Notice that $V_e$ and $V_h$ are such that the potential barriers push the electron and hole towards the MoS$_2$ and WS$_2$ regions, respectively, thus describing a type-II heterojunction profile\cite{lau2018interface, Bowing}. { We point out that, in principle, the formulas shown in Eq. (\ref{eq.bandoffset}), including a bowing parameter (i.e. a parabolic dependence on $\chi$) for the valence band, are developed for a large TMD flake with an uniform concentration $\chi$. However, it is clear that the conduction and valence band edges must exhibit a function profile that continuously interpolate their values in each side across the interface. Therefore, as a first approximation, we assume that, locally, the heterostructure potentials in each point of the interface follow the same functional dependence on $\chi$ as they would if they were in an equivalent alloy with concentration $\chi$. \cite{oliveira2004inhomogeneous,chaves2008grading, chaves2007excitonic}}

The Hamiltonian for the interacting electron-hole pair in this system takes the form
\begin{equation}\label{eq.Hamfull}
\begin{aligned}
    H = & -\frac{\hbar^2}{2}\frac{\partial}{\partial x_e} \frac{1}{m_e(x_e)}\frac{\partial}{\partial x_e} 
          -\frac{\hbar^2}{2 m_e(x_e)}\pdv[2]{}{y_e} \\
        & -\frac{\hbar^2}{2}\frac{\partial}{\partial x_h} \frac{1}{m_h(x_h)}\frac{\partial}{\partial x_h} 
      -\frac{\hbar^2}{2 m_h(x_h)}\pdv[2]{}{y_h}\\ 
        & + V_e(x_e) + V_h(x_h) + V_{eh}(|\vec{r_e}-\vec{r_h}|),
\end{aligned}
\end{equation}
where $m_e(x_e)$ and $m_h(x_h)$ are the effective masses of the electron and hole, respectively. These masses are assumed to depend linearly on the W concentration $\chi$ as $m_{e(h)}(x_{e(h)}) = m^W_{e(h)}\chi(x_{e(h)}) + m^{Mo}_{e(h)}[1-\chi(x_{e(h)})]$, so that the electron (hole) effective mass assumes its expected value for WS$_2$ in the left side of the system, $m^W_{e(h)} = 0.26 (0.34) m_0$, and changes linearly across the interface (namely, following the concentration profile) until it reaches its expected value for MoS$_2$ in the right side, $m^{Mo}_{e(h)} = 0.43 (0.53) m_0$, as illustrated in Fig. 2.


\begin{figure}[h!]
    \centering
    \includegraphics[width = \linewidth]{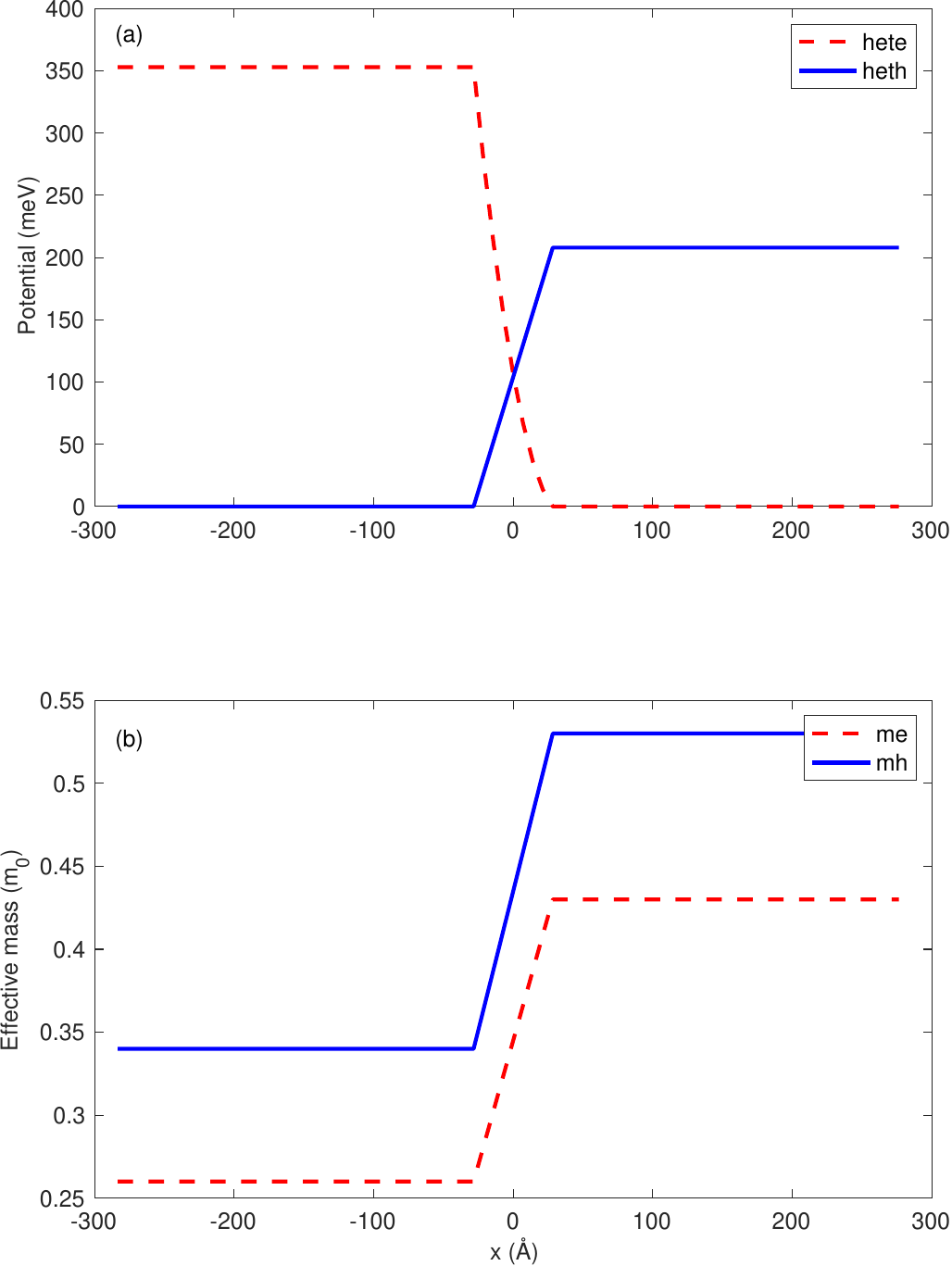}
    \caption{(color online) Sketch of (a) the potential and (b) the effective mass profiles for the electron (red dashed) and hole (blue solid) across the interface between WS$_2$ and MoS$_2$ within the side-by-side monolayer heterostructure.}
    \label{fig:System}
\end{figure}
Notice that the concept of effective masses originates from the periodic lattice of a crystal, which is not present in the interfacial region considered here, where the W concentration varies in space. However, it is customary to extrapolate this concept, as an approximation, also to the case of such smooth interfaces,\cite{stern1985calculated,shtinkov2000electronic,oliveira2004inhomogeneous,chaves2008grading} using the so-called Ben Daniel-Duke Hamiltonian.\cite{bendaniel1966space} Indeed, this is a plausible approximation, since it is clear that the effective masses in the left and right sides of the system sketched in Fig. \ref{fig:Sketch} are $m^W_{e(h)}$ and $m^{Mo}_{e(h)}$, respectively, and within a valid continuum model to describe this situation, it is reasonable to expect that these values would be smoothly interpolated across the interface. One natural way to interpolate these values is to assume that $m_{e(h)}$ depends linearly on $\chi$, as we did here, but a different dependence on $\chi$ can also be straightforwardly implemented in our model by adapting the $m_{e(h)}$ function in Eq. (\ref{eq.Hamfull}). \color{black} 

The $V_{eh}$ term in Eq. (\ref{eq.Hamfull}) is the e-h interaction potential. Since TMDs have similar values for dielectric constant, we assume a homogeneous dielectric constant $\epsilon$ = 14 $\epsilon_0$ along the interface, given by the average of dielectric constants of MoS$_2$ and WS$_2$. \cite{laturia2018dielectric} Similar approximation has been successfully employed in previous theoretical studies involving III-V semiconductor quantum wells. \cite{mathieu1992excitons,wang1999electric}
On the other hand, the dielectric constant changes dramatically along the out-of-plane direction, from the substrate to the TMD monolayer and the vacuum above it. Since the most common substrates, such as SiO$_2$, hexagonal boron-nitride (hBN), and sapphire, have dielectric constants smaller than that of the TMD, one can use the approximation proposed by Rytova \cite{rytova} and Keldysh \cite{keldysh} to obtain the e-h interaction potential
\begin{equation}
    V(r) = -\frac{\pi}{2(\epsilon_1 + \epsilon_2)\rho_0}\left[{H_0}\left(\frac{r}{\rho_0}\right) - {Y_0} \left(\frac{r}{\rho_0}\right)\right].
\end{equation}
where the screening length is $\rho_0 = 1 + 2 \pi \epsilon d/(\epsilon_1 + \epsilon_2)$, $d$ is the width of the TMD monolayer {(considered here as $d =$ 6.5\AA\,)} and $\epsilon_{1(2)}$ is the dielectric constant of the sub(super)strate. For practical purposes, we assume the substrate to be hBN, with $\epsilon_{1} = 4.5 \epsilon_{0}$, and different materials as superstrate.  

 \begin{figure}[h!]
    \centering
    \includegraphics[width = 0.9\linewidth]{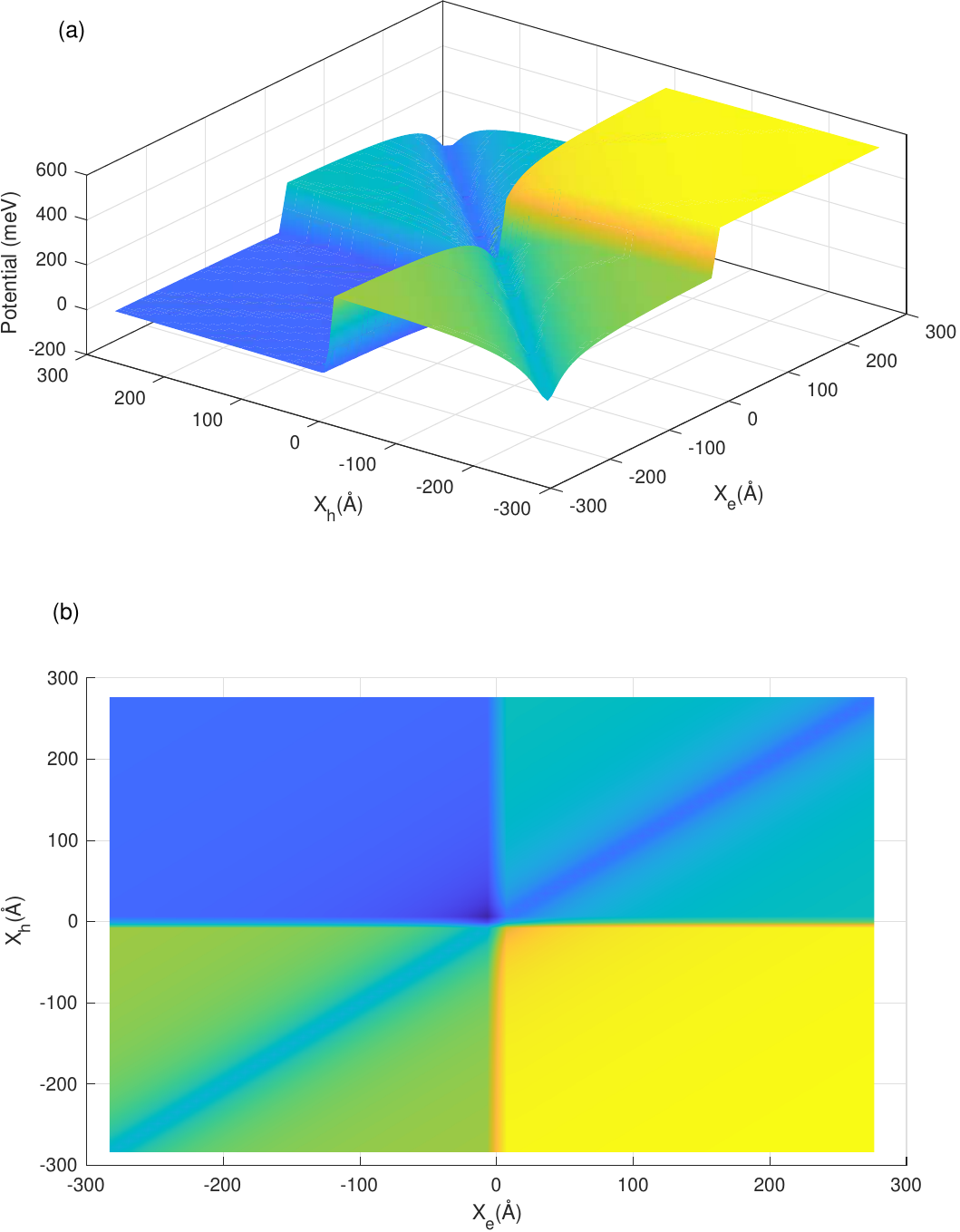}
    \caption{(color online) (a) Effective potential of the system, including both the e-h interaction, which is responsible for the dip along the $x_e = x_h$ line observed in the figure, and the heterostructure potential, which originates the steps observed in the potential. An interface width $w = 20$\AA\ and a dielectric constant $\epsilon_2 = 9.0\epsilon_0$ are assumed. (b) Color map of the effective potential shown in (a).}\color{black}
    \label{fig:pot_full}
\end{figure}

A competition between these potential contributions determines the exciton behavior. Without interactions, the lowest energy situation is that where electrons stay in the MoS$_2$ region (right), while holes stay in the WS$_2$ region (left) of the system sketched in Fig. \ref{fig:System}. However, the Rytova-Keldysh potential contribution will make electrons and holes pull each other towards the interfacial region where, provided the heterostructure potential barriers are high enough, they would form a type-II interface exciton. On the other hand, if the Rytova-Keldysh contribution is much stronger than the band offsets, particles may drag each other to the same material, in which case this electron-hole pair can no longer be seen as an interface exciton and is then a type-I exciton. {This is expected to be the case in monolayer TMDs under low dielectric screening by the environment, where the e-h binding energies of \cite{mueller2018exciton} $\approx$ 300 meV are in the same order of magnitude as the heterostructure potentials shown in Fig. 2(a).} The conditions leading to each of these exciton types in the MoS$_2$/WS$_2$ monolayer heterostructure will be discussed further on. 

We re-write Eq. (\ref{eq.Hamfull}) with $y = y_e - y_h$ being the relative coordinate in the $y$-direction, parallel to the interface:
\begin{equation}
    \begin{aligned}
        H = &-\frac{\hbar^2}{2}\frac{\partial}{\partial x_e} \frac{1}{m_e(x_e)}\frac{\partial}{\partial x_e} - \frac{\hbar^2}{2}\frac{\partial}{\partial x_h} \frac{1}{m_h(x_h)}\frac{\partial}{\partial x_h}\\+
            & V_e(x_e) + V_h(x_h) - \frac{\hbar^2}{2\mu(x_e,x_h)}\pdv[2]{}{y}\\
            + & V_{eh}(x_e,x_h,y).
    \end{aligned}
\end{equation}
As expected, the potential does not depend on the exciton center-of-mass along the $y$-direction, due to translational symmetry along this axis. As a consequence, the center-of-mass momentum along this direction is a good quantum number and is zero for the ground state. We propose a separation of variables in the form $\Psi(x_e,x_h,y) =\Psi_{eh}(x_e,x_h) \phi(y)$ in the Schr\"{o}dinger equation for the exciton, yielding
\begin{equation}\label{eq.final}
\begin{aligned}
    \left[-\frac{\hbar^2}{2}\frac{\partial}{\partial x_e} \frac{1}{m_e(x_e)}\frac{\partial}{\partial x_e} - \frac{\hbar^2}{2}\frac{\partial}{\partial x_h} \frac{1}{m_h(x_h)}\frac{\partial}{\partial x_h}+
        V_e(x_e)\right.\\ \left. + V_h(x_h)\right]\Psi_{eh} 
     +\overline{V}_{eh}(x_e,x_h) \Psi_{eh} = E\Psi_{eh}, 
\end{aligned}
\end{equation}
where the effective interaction potential is given by
\begin{equation}
    \begin{aligned}
    \overline{V}_{eh}(x_e,x_h) 
    =
    \int V_{eh}(x_e,x_h,y)\phi^2dy
    -\\
    \frac{\hbar}{2\mu(x_e,x_h)}\int \phi\pdv[2]{}{y}\phi dy.
    \end{aligned}
\end{equation}

The effective form of the potential, shown in Fig. \ref{fig:pot_full}, is defined as $V_{Effective} = V_e(x_e) + V_h(x_h) + \bar{V}_{eh}(x_e,x_h)$, which includes both the e-h interaction and heterostructure contributions. An interface width $w = 20$\AA\ and a superstrate dielectric constant $\epsilon_2 = 9.0\epsilon_0$ are assumed in Fig. \ref{fig:pot_full}.

Within the variational principle, we use a hydrogenic trial function $\phi(y) = A\exp({-|y|}/{a})$, with a variational parameter $a$, so that
\begin{equation}\label{eq.veh}
    \begin{aligned}
    \overline{V}_{eh}(x_e,x_h,a) 
    =
    \frac{\hbar^2}{2\mu(x_e,x_h) a^2} + \\
    \frac{1}{a} \int_{-\infty}^{+\infty}V_{eh}(x_e,x_h,y)e^{-2|y|/a}dy.
    \end{aligned}
\end{equation}
where the normalization parameter is $A = 1/\sqrt{a}$.

Finally, the two-variable differential equation, Eq. (\ref{eq.final}), is numerically solved with a finite difference scheme and numerical diagonalization, {see Appendix}. The exciton energy is thus obtained by minimization of the energy $E(a)$, i.e. $E_{exc} = min\{E(a)\}$.

Once the wave function is obtained by the numerical procedures and minimization process, we calculate the e-h overlap as 
\begin{equation}\label{eq.overlap}
 O_{eh}= \int_{-\infty}^{\infty} \int_{-\infty}^{\infty}|\Psi(x_e,x_h)|^2 \delta(x_e-x_h) dx_e dx_h|\phi(0)|^2,    
\end{equation}
which allows us to estimate the oscillator strength of the exciton state. In what follows, we discuss how the exciton energy and overlap, as calculated with this method, are affected by the interface width and dielectric environment.

\section{Results and discussion}

\begin{figure}[h!]
    \centering
    \includegraphics[width = \linewidth]{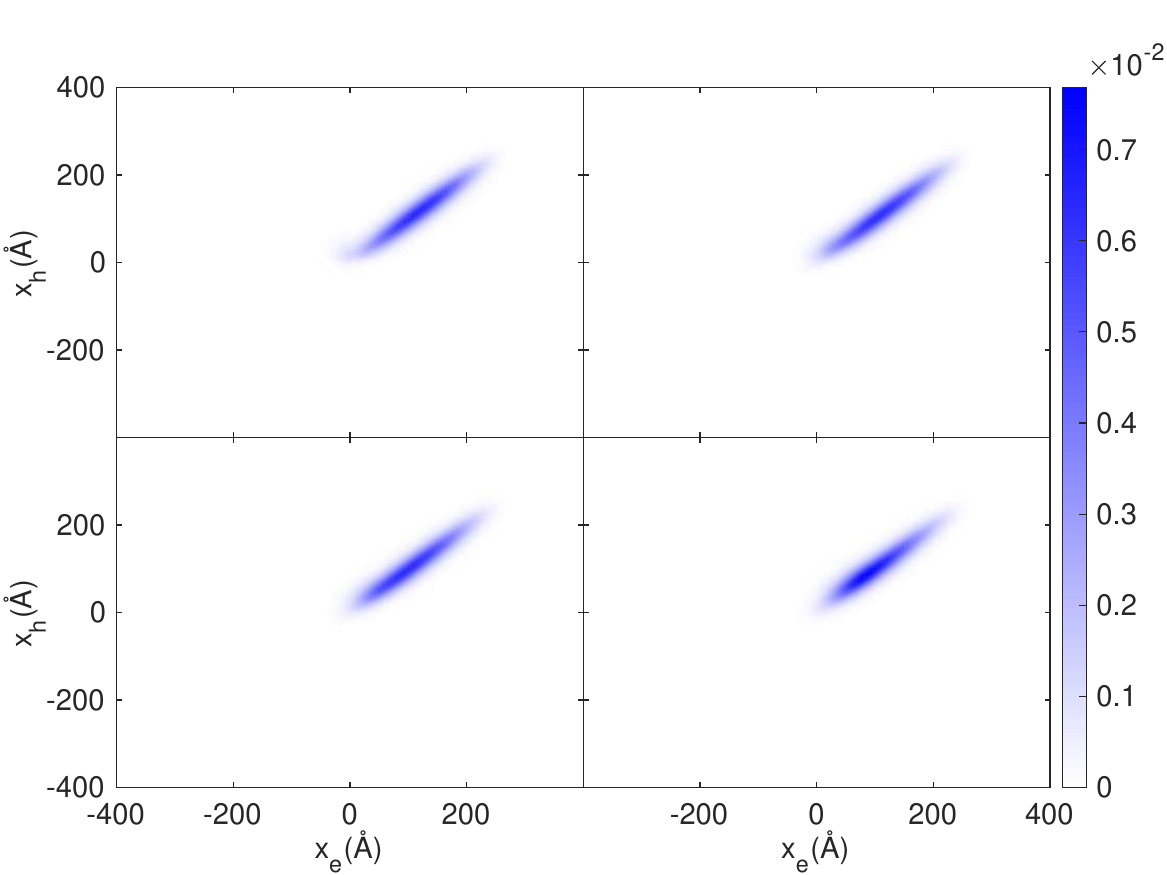}
    \caption{(Color online) Color maps of the probability density distribution assuming an environment with dielectric constant $\epsilon_2 = 1.0\epsilon_0$ and interface widths (a) $w = 10$ \AA\,, (b) $25$ \AA\,, (c) $40$ \AA\,, and (d) $60$ \AA\,.}
    \label{fig:WF_1}
\end{figure}

\begin{figure}[h!]
    \centering
    \includegraphics[width = \linewidth]{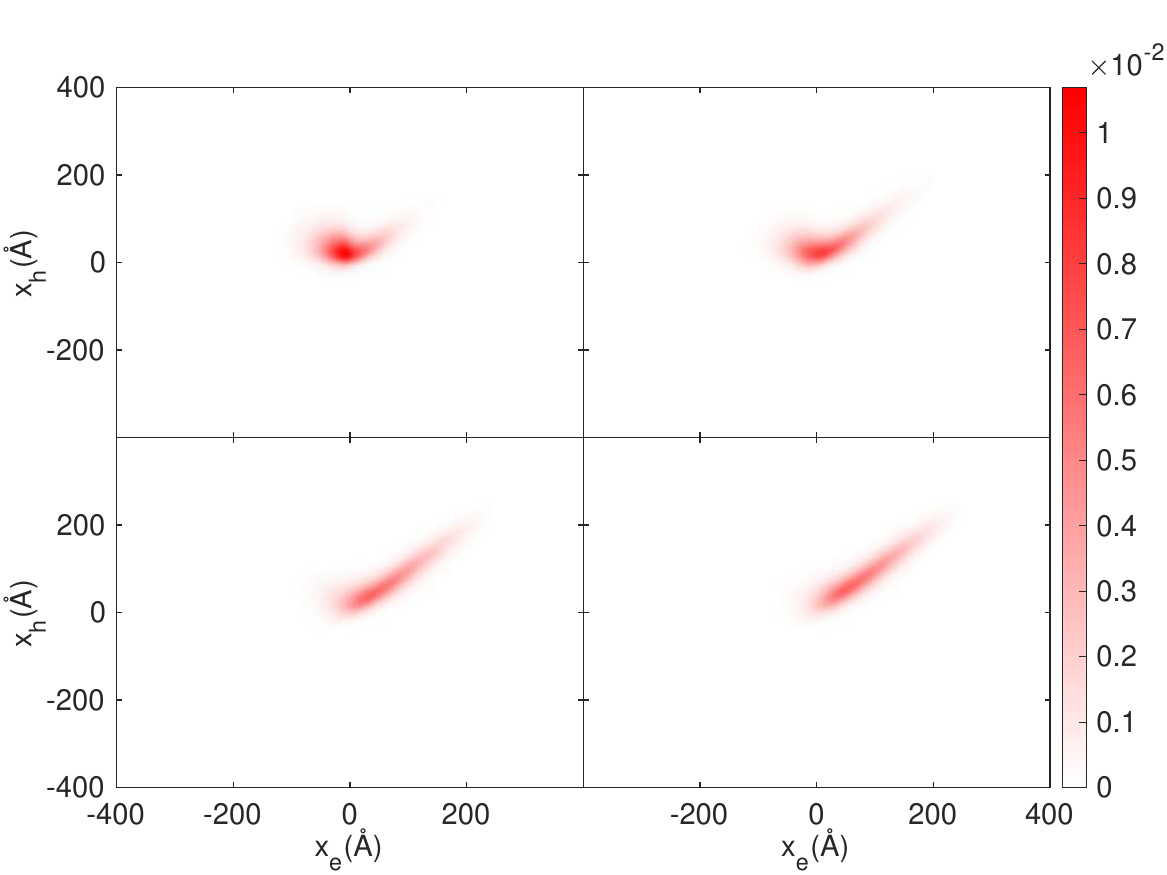}
    \caption{(Color online) Same as in Fig. \ref{fig:WF_1}, but assuming $\epsilon_2$ = 4.5 $\epsilon_0$.}
    \label{fig:WF_4nh}
\end{figure}

\begin{figure}[h!]
    \centering
    \includegraphics[width = \linewidth]{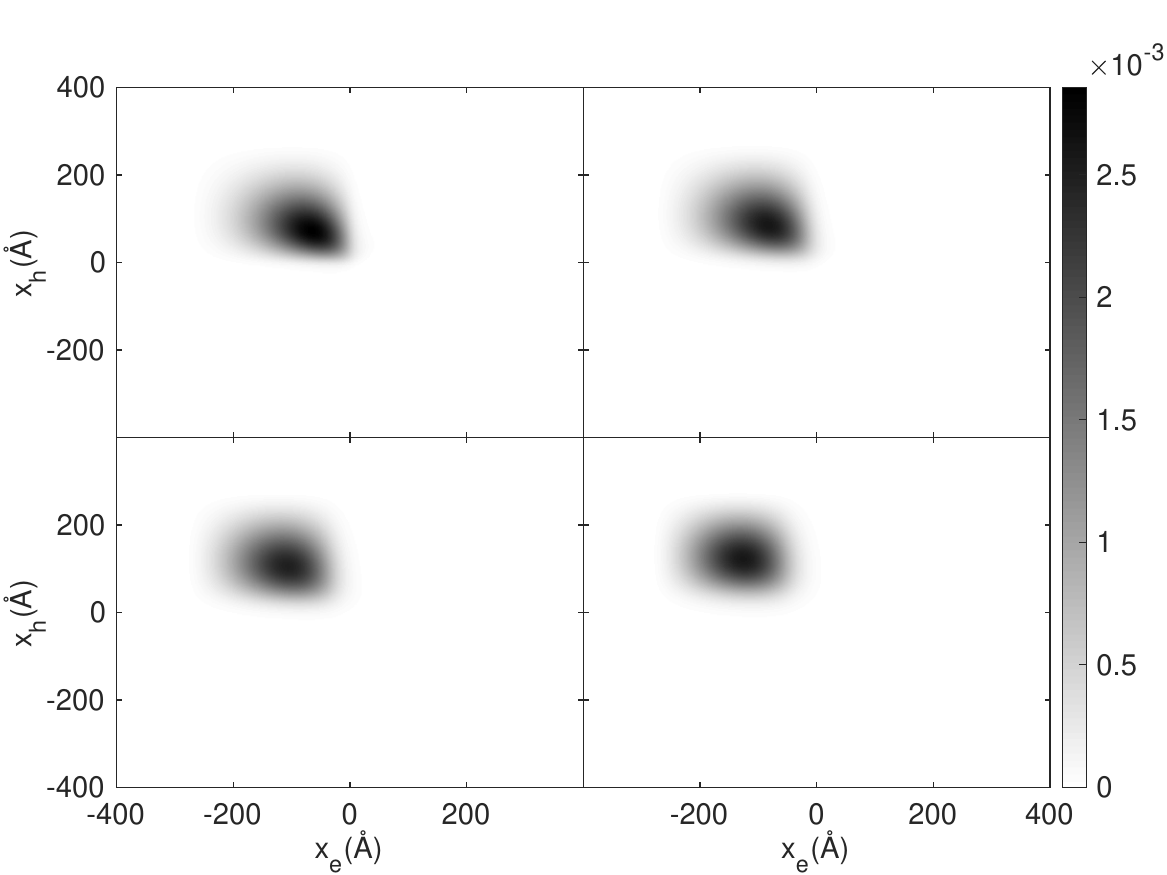}
    \caption{(Color online) Same as in Fig. \ref{fig:WF_1}, but assuming $\epsilon_2$ = 9 $\epsilon_0$.}
    \label{fig:WF_6}
\end{figure}

Figures \ref{fig:WF_1}, \ref{fig:WF_4nh} and \ref{fig:WF_6} show the probability density functions $|\Psi(x_e,x_h)|^2$ assuming different dielectric constants $\epsilon_2$ = 1 $\epsilon_0$, 4.5 $\epsilon_0$, and 9 $\epsilon_0$, respectively. In each figure there are four different panels, for interface widths (a) 10 \AA\,, (b) 25 \AA\, (c) 40 \AA\, and (d) 60 \AA\,.

As previously mentioned, the band offsets in this combination of TMDs is such that a type-II exciton is expected. However, in Fig. \ref{fig:WF_1}, the exciton wave function has its maxima along the $x_e = x_h$ line, extended to the $x_e < 0$, $x_h < 0$ region, for all values of interface width considered here. This means that electron and hole, although bound to each other as an exciton, are both at the MoS$_2$ region. Indeed, due to the low dielectric screening by the environment, the Rytova-Keldysh contribution to the potential in this case is sufficient to overcome the heterostructure potential and, therefore, this state can not be seen as an interface (type-II) exciton, but rather as a type-I exciton at MoS$_2$. Notice that the wave function goes to zero at the edge of our computational box at $x_e = x_h = - 400$ \AA\,, as a result of a spurious confinement caused by the limited size of the box. However, we carefully verified that we have chosen a computational box large enough as to produce an error of at most $ 1$ meV in the exciton energies.

In Fig. \ref{fig:WF_4nh}(a), on the other hand, the probability density peaks in the surroundings of $x_e = x_h = 0$, with a very small dispersion away from this point. This is consistent with a bound e-h pair where both charge carriers are close to the interface, thus characterizing an actual interface exciton state. In this case, the dielectric screening caused by the hBN substrate/superstrate suppresses the e-h interaction which now, although still strong, is not strong enough to overcome the bands mismatch and pull particles away from the region where their corresponding band edges are more energetically favorable. However, as the interface width increases, the heterostructure potential effectively weakens and, consequently, the situation becomes similar to that of Fig. \ref{fig:WF_1}, with electron and hole both at MoS$_2$. 


As for the wave functions shown in Fig. \ref{fig:WF_6}, where  a higher dielectric constant is assumed for the superstrate material (namely, $\epsilon_2$ = 9 $\epsilon_0$), they are compatible with a state that can be seen as a type-II  exciton: the wave function is on the up-left quadrant of Fig. \ref{fig:WF_6}, which means electrons (holes) are mostly in a region of negative (positive) $x$-coordinate, namely, in the Mo(W)S$_2$ material, although still being pushed towards the interface due to the attractive interaction with the other charge. As the width of the interface region increases, from Fig. \ref{fig:WF_6}(a) to Fig. \ref{fig:WF_6}(d), the wave function becomes less confined at the interface and rather spreads more over the up-left quadrant.

\begin{figure}[h!]
    \centering
    \includegraphics[width = \linewidth]{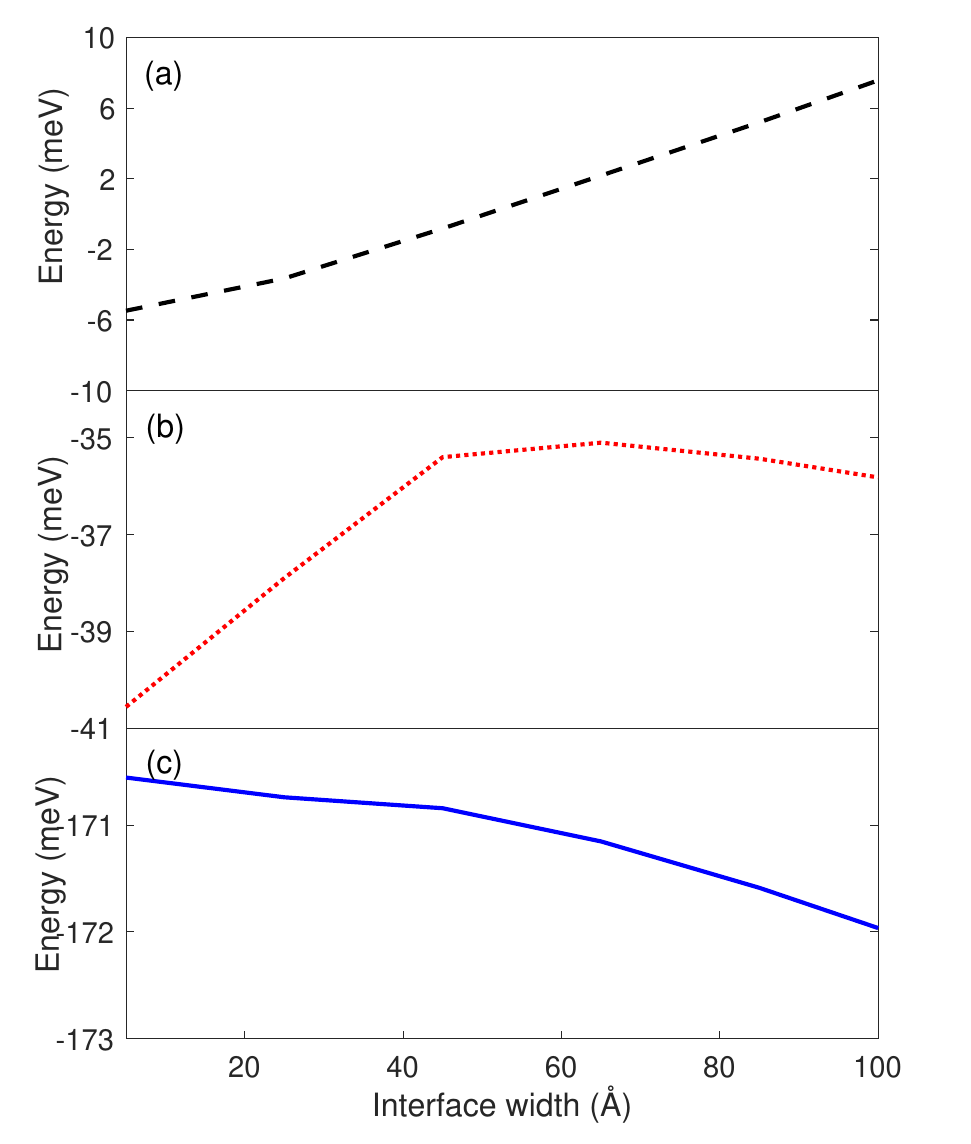}
    \caption{(color online) Dependence of the exciton energy on the interface width, for (a) $\epsilon_2 = 9.0 \epsilon_0$, (b) $\epsilon_2 = 4.5 \epsilon_0$, and $\epsilon_2 = 1.0 \epsilon_0$. Notice that, in this reference frame, lower energies represent excitons with higher (more negative) binding energy. \color{black}}
    \label{fig:Energy}
\end{figure}

The results in Figs. \ref{fig:WF_1} to \ref{fig:WF_6} are better understood by analyzing the exciton energies in each case, which are shown in Fig. \ref{fig:Energy} as a function of the interface width $w$. In the case of a sapphire substrate, the fact that we have a type-II interface exciton for all values of interface width considered here is reflected in its exciton energy in Fig. \ref{fig:Energy}(a), which only increases with $w$, since the longer e-h separation for such a type-II exciton in wider interfaces yield weaker e-h binding energies. For a superstrate with lower dielectric constant such as hBN ($\epsilon_2$ = 4.5 $\epsilon_0$), the exciton energy in Fig. \ref{fig:Energy}(b) is demonstrated to be lower as compared to the one of the type-II exciton in sapphire ($\epsilon_2$ = 9 $\epsilon_0$). In this case, as the interface width $w$ increases, the energy increases until it it peaks at $w \approx 60$ \AA\,, when the exciton can no longer be seen as an interface exciton and is rather a MoS$_2$ exciton. It is yet surprising that the energy of such a type-I MoS$_2$ exciton in Fig. \ref{fig:Energy} seem lower than those observed for similar excitons in monolayer MoS$_2$ without an interface with WS$_2$, even in the presence of hBN as surrounding material, where exciton binding energies are still expected to be of a few hundreds of meV. The reason for such apparently lower binding energy in Fig. \ref{fig:Energy} lies in the fact that the energy of such type-I MoS$_2$ exciton state, obtained here by solving Eq. (\ref{eq.final}), whose energy reference frame is based on the bands mismatch in the hetero junction, is actually the exciton binding energy plus the band offset (of 208 meV, see Eq. (\ref{eq.bandoffset})) for the hole, which, in the absence of the Rytova-Keldysh interaction, was supposed to be rather at WS$_2$, where its valence band energy is lower. In the absence of a superstrate material ($\epsilon_2$ = $\epsilon_0$), binding energies in Fig. \ref{fig:Energy}(c) are shown to be stronger and less susceptible to the interface width, since the exciton is already in a type-I state fully confined at the MoS$_2$ side of the heterostructure. 

\begin{figure}[h!]
    \centering
    \includegraphics[width = \linewidth]{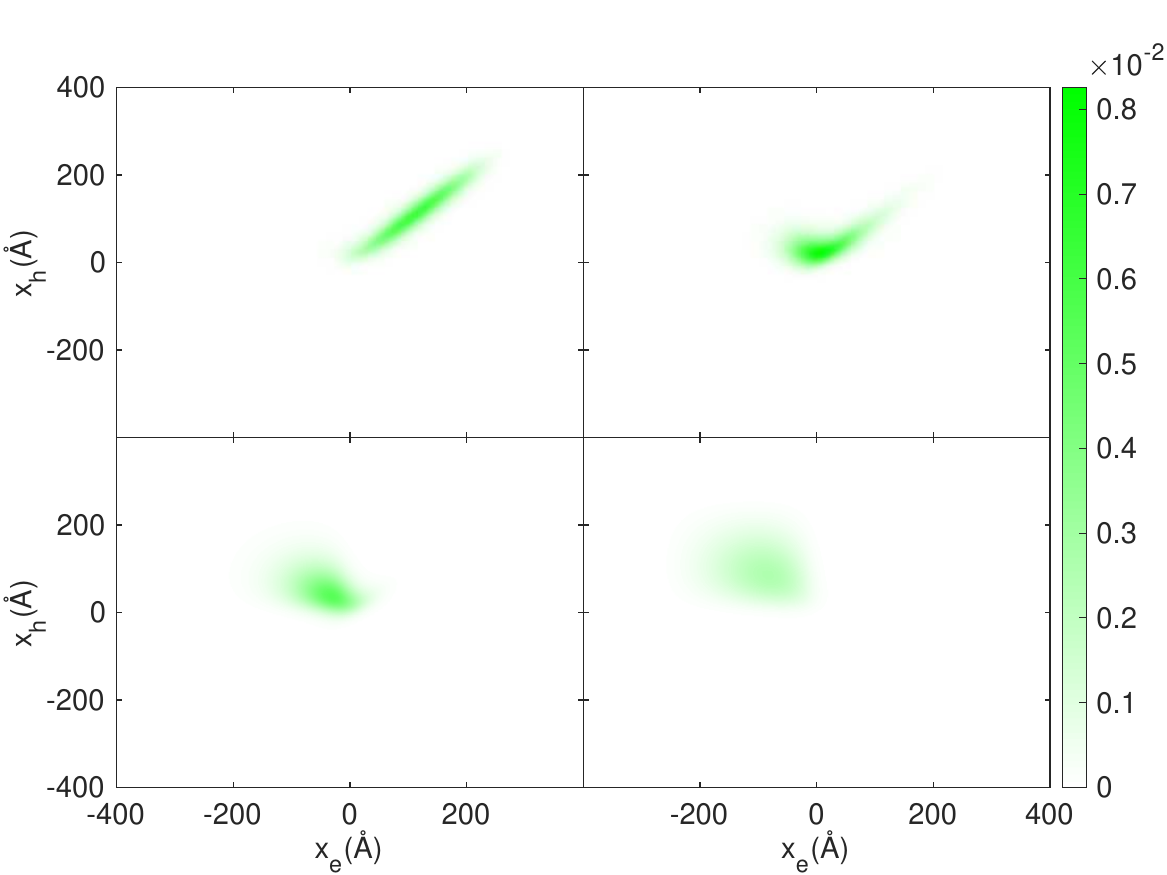}
    \caption{(Color online) $|\Psi|^2$ for a fixed interface width of $10$ \AA\, assuming (a) $\epsilon_2$ = 1$\epsilon_0$, (b) $\epsilon_2$ = 3$\epsilon_0$, 
    (c) $\epsilon_2$ = 5$\epsilon_0$, 
    (d) $\epsilon_2$ = 9$\epsilon_0$.}
    \label{fig:WF_eps}
\end{figure}

The probability densities shown in Fig. \ref{fig:WF_eps} summarize the transition from an effectively type-I exciton, when the superstrate is vacuum ($\epsilon_2 = 1 \epsilon_0$), see panel (a), to an effectively spatially separated exciton, when $\epsilon_2 = 9 \epsilon_0$, see panel (d). In fact, we have numerically verified that using materials with dielectric constant higher than $\approx 5 \epsilon_0$, such as sapphire and MgO, as superstrate, we expected to observe effectively type-II excitons in MoS$_2$/WS$_2$ planar heterostructures. \cite{borghardt2017engineering}  

\begin{figure}[h!]
    \centering
    \includegraphics[width = \linewidth]{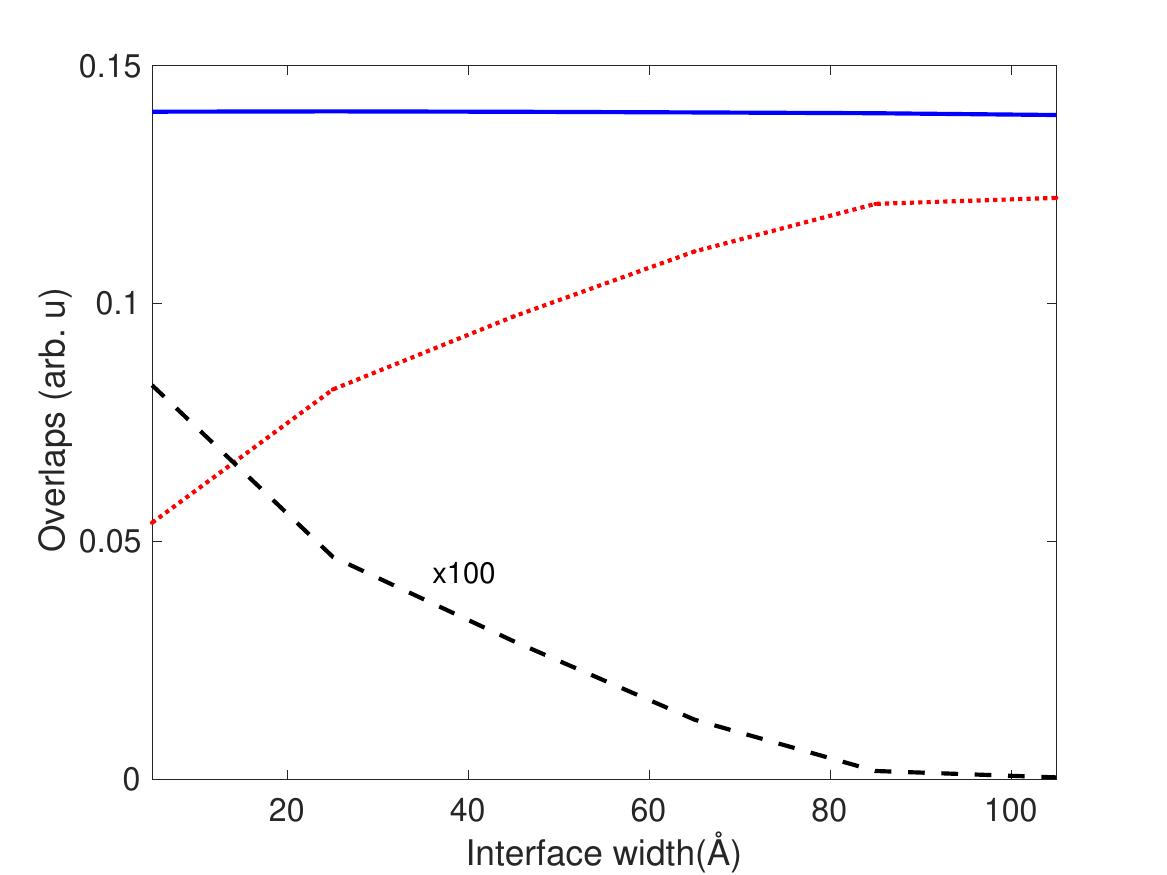}
    \caption{(color online) Dependence of the e-h overlap  on the interface width for $\epsilon_2 = 1.0$ (blue solid), $\epsilon_2 = 4.5$ (red dotted), $\epsilon_2 = 9.0$ (back dashed).}
    \label{fig:Overlaps}
\end{figure}

Finally, {Fig. \ref{fig:Overlaps} shows the e-h overlap $O_{eh}$ as calculated by Eq. (\ref{eq.overlap}). Within the Elliot theory of excitonic transitions\cite{haug2009quantum}, the oscillator strength for excitonic processes is proportional to $|\psi(r=0)|^2$, namely, to the probability density function of the exciton relative motion at the point where the e-h relative distance is zero. This quantity is equivalent to the e-h overlap  $O_{eh}$ calculated here, therefore, higher (lower) values of $O_{eh}$ would lead to higher (lower) oscillator strengths and, consequently, to higher (lower) transition probabilities and absorption peaks in the reflectance spectra of these systems. Likewise, for light emission and photoluminescence, the behavior of $O_{eh}$ provides a qualitative view of the emission rate and exciton lifetime: smaller oscillator strength represent longer exciton lifetimes. \cite{van1987giant}} 

The e-h overlap in Fig. \ref{fig:Overlaps} for the case of $\epsilon_2 = 1.0\epsilon_0$ (blue solid) is the highest and it is practically constant over any value of interface width $w$, which is consistent with the fact that this is not a charge separated exciton state, regardless of the interface width. Conversely, for $\epsilon_2$ = 4.5 $\epsilon_0$ (red dotted), the smaller overlap for narrow interface width indicates an interface exciton, although the overlap increases towards the value for a non-separated exciton as the interface is made wider. As for the $\epsilon_2$ = 9.0 $\epsilon_0$ case (black dashed), the overlap is reduced by two orders of magnitude, suggesting an interface exciton with strong charge separation, whose overlap actually \textit{decreases} with increasing interface width, in contrast with the other two cases shown in the figure. Although the smaller overlap in the $\epsilon_2$ = 4.5 $\epsilon_0$ and 9.0 $\epsilon_0$ cases would hinder the experimental observation of these excitons with light absorption techniques, they are still expected to be observed in photoluminescence experiments (just like e.g. type-II inter-layer excitons in van der Waals stacked heterostructures\cite{rivera2015observation}) and they are ideal for applications where long exciton lifetimes are needed. 

\section{Conclusions}

In summary, we have theoretically investigated the energy and electron-hole overlap of excitons at an interface between two monolayer TMDs arranged as a side-by-side heterostructure, within the effective mass approximation and Wannier-Mott theory. MoS$_2$ and WS$_2$ are chosen as heterostructure materials, for practical purposes. The interface between the TMDs is allowed to be smooth in our model, by assuming a W$_{\chi}$Mo$_{1-\chi}$S$_2$ alloy at the interface, whose W concentration $\chi$ increases linearly from 0 (MoS$_2$) to 1 (WS$_2$) along the interface.  We consider a monolayer heterostructure on top of a hBN substrate and assume either vaccum, hBN or sapphire as possible capping environments. Our results demonstrate that under vaccuum, interface excitons, as defined by charge separated e-h pairs, are not energetically favorable, since the strong e-h attraction overcomes the heterostructure potential that would push charge carriers towards different materials. A hBN cover weakens, via dielectric screening, the e-h interaction in the monolayer heterostructure, which allows for the existence of an interface exciton as the lowest energy solution, provided the interface between materials is sharp. However, such an interface exciton does not survive as the interface between the two materials is made wider. A sapphire capping environment, on the other hand, provides enough screening to guarantee a pure interface exciton as the lowest energy solution for any value of interface width considered here. In fact, in the latter case, increasing the interface width reduces even further the e-h overlap, yielding even longer interface exciton lifetimes. 

The absence of interface excitons for the case of heterostructures under vacuum, or even for heterostructures with non-abrupt interfaces under hBN, as verified by our calculations, sheds light on the fact that the experimental observation of interface excitons in side-by-side heterostructures, which are usually under these dielectric environments in actual experiments, has been elusive so far. Our results suggest that a capping material that provides higher dielectric screening is essential for the existence of interface excitons, even though the naturally lower e-h overlap of such charge separated excitons would still hinder its observation with light absorption techniques.

We believe these results will guide future researchers towards the experimental visualization of interface excitons in side-by-side van der Waals heterostructures, as well as the design of optoelectronic devices and applications that require the long exciton lifetimes expected for such charge separated excitons, e.g. photodetectors and solar cells, within monolayer atomic scale.

\acknowledgements This work was financially supported by the Brazilian Council for Research (CNPq), through the PQ and UNIVERSAL programs, and by the Research Foundation - Flanders (FWO-Vl).

\vspace{0.5 cm}
\section*{Apendix: Discretized Hamiltonian and numerical method}

The Schrodinger-like effective mass equation for the Hamiltonian Eq. (\ref{eq.final}) is a two-dimensional differential equation in the coordinates $x_e$ and $x_h$, which is solved here by a finite difference scheme and numerical diagonalization of the resulting Hamiltonian matrix. In order to do so, the $x_{e(h)}$-coordinate is discretized into a $N = 120$ point grid (thus forming a $N \times N$ two-dimensional computational mesh) with step length $\Delta x = $ \AA\, and associated to the index $i(j)$, so that the kinetic energy terms are re-written as
\begin{eqnarray}
 -\frac{\hbar^2}{2} \frac{\partial}{\partial x_{e(h)}}\frac{1}{\mu(x_e,x_h)}\frac{\partial}{\partial x_{e(h)}}\Psi(x_e,x_h) \approx \nonumber\\
A^{e(h)}_{i,j}\Psi_{i+1(i),j(j+1)} + B^{e(h)}_{i,j}\Psi_{i,j}+C^{e(h)}_{i,j}\Psi_{i-1(i),j(j-1)},
\end{eqnarray}
with factors given by
\begin{eqnarray}
A^{e(h)}_{i,j} = -\frac{\hbar^2}{4\Delta x^2}\left[\frac{1}{\mu_{i+1,j(i,j+1)}} + \frac{1}{\mu_{i,j}}\right] \\ 
B^{e(h)}_{i,j} = \frac{\hbar^2}{8\Delta x^2}\left[\frac{1}{\mu_{i+1,j(i,j+1)}} + \frac{2}{\mu_{i,j}} + \frac{1}{\mu_{i-1,j(i,j-1)}}\right]\\
C^{e(h)}_{i,j} = -\frac{\hbar^2}{4\Delta x^2}\left[\frac{1}{\mu_{i,j}} + \frac{1}{\mu_{i-1,j(i,j-1)}}\right].
\end{eqnarray}
The discretized effective potential function ${V}_{Effective}^{i,j}$ is calculated by numerical integration of the expression in Fig. \ref{eq.veh} for a given value of the variational parameter $a$ and then added to the diagonal terms $B^e_{i,j} + B^h_{i,j}$. The resulting Hamiltonian is represented by a block five-diagonal matrix, \cite{chaves2015split} which is then diagonalized by numerical procedures. The process is repeated for several values of the variational parameter $a$ until an energy minimum is found.



%
\end{document}